\documentclass[aps,pra,twocolumn,english,superscriptaddress,showpacs,showkeys,nofootonbib]{revtex4}

\usepackage{amssymb}
\usepackage{amsmath}
\usepackage{amsthm}
\usepackage{amsfonts}
\usepackage{color}
\usepackage{amscd}
\usepackage{graphicx}

\newtheorem{thm}{Theorem}[section]

\newtheorem{defin}{Definition}[section]
\newtheorem{prop}{Proposition}[section]

\newtheorem{corol}{Corollary}[section]
\newcommand{\ket}[1]{|#1\rangle}
\newcommand{\bra}[1]{\langle #1 |}

\newcommand{\nn}{\nonumber}

\newcommand{\rg}{\mathop{\rm r }\nolimits}
\newcommand{\lin}{\mathop{\rm span }\nolimits}
\makeatother
\begin{document}

\title{Inductive classification of multipartite entanglement under SLOCC}

\author{L. Lamata}
\email[]{lamata@imaff.cfmac.csic.es} \affiliation{Instituto de
Matem\'{a}ticas y F\'{\i}sica Fundamental, CSIC, Serrano 113-bis,
28006 Madrid, Spain}

\author{J. Le\'{o}n}
\email[]{leon@imaff.cfmac.csic.es} \affiliation{Instituto de
Matem\'{a}ticas y F\'{\i}sica Fundamental, CSIC, Serrano 113-bis,
28006 Madrid, Spain}

\author{D. Salgado}
\email[]{david.salgado@uam.es} \affiliation{Dpto.\ F\'{\i}sica
Te\'{o}rica, Universidad Aut\'{o}noma de Madrid, 28049 Cantoblanco,
Madrid, Spain}
\author{E. Solano\footnote{Present address: Physics
Department, ASC, and CeNS, Ludwig-Maximilians-Universit\"at,
Theresienstrasse 37, 80333 Munich, Germany}}
\email[]{enrique.solano@physik.lmu.de}
\affiliation{Max-Planck-Institut f\"{u}r Quantenoptik,
Hans-Kopfermann-Strasse 1, 85748 Garching, Germany}
\affiliation{Secci\'{o}n F\'{\i}sica, Departamento de Ciencias,
Pontificia Universidad Cat\'{o}lica del Per\'{u}, Apartado Postal
1761, Lima, Peru}
\date{\today}

\begin{abstract}
We propose an inductive procedure to classify $N-$partite entanglement under stochastic local operations and classical communication (SLOCC) provided such a classification is known for $N-1$ qubits. The method is based upon the analysis of the coefficient matrix of the state in an arbitrary product basis. We illustrate this approach in detail with the well-known bi- and tripartite systems, obtaining as a by-product a systematic criterion to establish the entanglement class of a given pure state without resourcing to any entanglement measure. The general case is proved by induction, allowing us to find an upper bound for the number of $N$-partite entanglement classes in terms of the number of entanglement classes for $N-1$ qubits.
\end{abstract}
\maketitle
\section{Introduction}
\label{Intro}
Entanglement resides in the root of the most surprising quantum phenomena (cf.\ e.g.\ \cite{Per93a}). Furthermore, it is the main resource in the usage of quantum systems to process information \cite{BenDiV00a} in tasks such as cryptographic key distribution \cite{DusLutHen06a}, quantum computation \cite{DeuEke98a,RauBri01a}, quantum state teleportation \cite{BouEkeZei00a}, quantum communication \cite{HorHorHor01a} and dense coding \cite{BenWie92a}. However a comprehensive understanding of entanglement is still lacking, mainly because it is a highly counterintuitive feature of quantum systems (non-separability \cite{Bel87a}) and because its analysis can be undertaken under different, although complementary, standpoints \cite{EisGro05a}. As prominent examples the subjects of deciding in full generality whether a given state carries entanglement or not and how much entanglement the system should be attributed to are vivid open questions (cf.\ e.g.\ \cite{Bru02a} and references therein). This state of affairs is critical in multipartite systems, where most applications find their desired utility.\\
Among others, part of the efforts are being dedicated to classify  under diversely motivated criteria the types of entanglement which a multipartite system can show. It is in this sense desirable, independently of these criteria, to have classification methods valid for any number $N$ of entangled systems. One of these most celebrated criteria to carry out such a classification was provided in \cite{DurVidCir00a}. In physical terms D\"{u}r \emph{et al.} defined an entanglement class as the set of pure states which can be interrelated through stochastic local operations and classical communications (SLOCC hereafter) or equivalently, as those pure states which can carry out the same quantum-informational tasks with non-null possibly different probabilities. They also proved the mathematical counterpart of this characterization: two states $\Psi$ and $\bar{\Psi}$ of a given system belong to the same entanglement class if, and only if, there exist invertible local operators (ILO's hereafter; that is, nonsingular matrices), which we agree on denoting as $F^{[i]}$ such that $\bar{\Psi}=F^{[1]}\otimes\dots\otimes F^{[N]}(\Psi)$. Moreover, they provided the first classification under this criterion of tripartite multiqubit entanglement, giving birth to the two well-known genuine entanglement triqubit classes named as $GHZ$ and $W$ classes. Later on, exploiting some accidental facts in group theory, Verstraete \emph{et al.} \cite{VerDehMooVer02a} gave rise to the classification of $4$-qubit states.\\
 Regretfully none of the previous works allowed one to succeed in obtaining a generalizable method. In the second case, the exploitation of a singular fact such as the isomorphism $SU(2)\otimes SU(2)\simeq SO(4)$ is clearly useless in a general setting; in the first case, the use of quantitative entanglement measures specifically designed for three qubits, as the $3-$tangle \cite{CofKunWoo00a}, to discern among different entanglement classes discourages one to follow up the same trend, since we would have to be able to build more generic entanglement measures, \emph{per se} a formidable task. However, Verstraete \emph{et al.} \cite{VerDehMoo03a} succeeded in this approach by introducing the so-called normal forms, namely those pure states such that all reduced local operators are proportional to the identity matrix. These authors also provided a systematic, mostly numerical, constructive procedure to find the ILO's bringing an arbitrary pure state to a normal form. Furthermore, the use of these normal forms allowed them to introduce entanglement measures (entanglement monotones \cite{Vid00a}, indeed), which offered the possibility to quantify the amount of entanglement in the original state. In this same trend, other alternatives can also be found in the literature \cite{OstSie05a,LovMaaSmiAmiGraIliIzmZag06a,RigOliOli06a}. For completeness' sake let us recall that classification under SLOCC is coarser than that using only local unitaries, that is in which every $F^{[k]}$ is unitary. Nevertheless relevant results in this realm can also be found in the literature \cite{GraRotBet98a,AciAndCosJanLatTar00a,CarHigSud02a,GaoAlbFeiWan06a}.\\
Here we offer an alternative and complementary approach to the classification under SLOCC based on an analysis of the singular value decomposition (SVD) of the coefficient matrix of the pure state in an arbitrary product basis. The coefficient matrix is chosen according to the partition $1|2\dots N$ with the subsequent goal of establishing a recursive procedure allowing one to elucidate the entanglement classes under SLOCC provided such a classification is known with one less qubit. The key feature in this scheme is the structure of the right singular subspace, i.e.\ of the subspace generated by the right singular vectors of the coefficient matrix, set up according to the entanglement classes which its generators belong to. As a secondary long-term goal, the approach seeks possible connections to the matrix product state (MPS) formalism (cf.\ e.g.\ \cite{Eck05a} and multiple references therein), which is becoming increasingly ubiquitous in different fields such as spin chains \cite{AffKenLieTas87a}, classical simulations of quantum entangled systems \cite{Vid03a}, density-matrix renormalization group techniques \cite{VerPorCir04a} and sequential generation of entangled multiqubit states \cite{SchSolVerCirWol05a}.\\
 We have preferred the mathematical conventions. The canonical orthonormal basis in $\mathbb{C}^{N}$ will be denoted by $\{e_{j}\}_{j=1,\dots,N}$ (correspondingly in physics the kets $|j-1\rangle$). Normalization is not relevant in elucidating the entanglement class which a state belongs to. Thus we will deal with unnormalized vectors and non-unit-determinant ILO's. In the SVD of an arbitrary matrix (cf.\ appendix), $V$ and $W$ will denote the left and right unitary matrix, whereas $\Sigma$ will stand for the diagonal possibly rectangular matrix with the singular values as entries. In the multiqubit cases, we will agree on denoting by small Greek letters $\phi,\varphi,...$ vectors belonging to $\mathbb{C}^{2}$, whereas capital Greek letters $\Psi,\Phi,...$ will denote a generic entangled state in $\mathbb{C}^{2}\otimes\mathbb{C}^{2}$.\\
The paper is organized as follows. In section \ref{BipEnt} the entanglement of two qubits is revisited with a reformulation of the Schmidt decomposition criterion in terms of the singular subspaces. In section \ref{TriEntSec} the extension to the three-qubit case is developed in detail and the principles of the generalization to multipartite and arbitrary-dimension systems are discussed in section \ref{Gener}. We close with some concluding remarks in section \ref{Con}. An appendix with the relevant facts about the SVD is also included.
\section{Bipartite entanglement}
\label{BipEnt}
\subsection{The Schmidt decomposition criterion revisited}
\label{SchDecRev}
The determination of entanglement of pure states of bipartite systems in any dimensions, in general, and in two dimensions (qubits), in particular, was solved long ago with the aid of the well-known Schmidt decomposition \cite{Sch07a,EkeKni95a}, by which any bipartite state can be written as a biorthogonal combination
\begin{equation}
\Psi=\sum_{n=1}^{\min(N_{1},N_{2})}\lambda_{n}\phi_{n}^{(1)}\otimes\psi_{n}^{(2)},
\end{equation}
\noindent where $\lambda_{1}\geq\lambda_{2}\geq\dots\geq 0$ for all $n$ and $N_{i}$ denotes the dimension of subsystem $i$. If $\lambda_{n}=0$ except for only one index $\lambda_{1}\neq 0$, then the state is a product state; on the contrary, if $\lambda_{n}\neq 0$ for two or more indices, then the state is an entangled state. Furthermore, $\lambda_{n}^{2}$ coincides with the common eigenvalues of both reduced density operators. Thus, to practically determine the entangled or separable character of a given pure state all we must do is to compute the spectrum of $\rho_{1}$ or $\rho_{2}$ or equivalently to analyze the dimensionality of their ranges. This is the backbone in the study of $3-$partite entanglement  carried out in \cite{DurVidCir00a}.\\
Followingly in order to pave the way for a generalization to multipartite systems, we will reformulate the Schmidt decomposition criterion for bipartite systems focusing upon the subspace generated by the singular vectors. We need the next
\begin{defin}
We will denote by $\mathfrak{V}$ (resp.\ $\mathfrak{W}$) the subspace generated by the left (resp.\ right) singular vectors, i.e.\ $\mathfrak{V}=\lin\{v_{1},\dots,v_{k}\}$ (resp. $\mathfrak{W}=\lin\{w_{1},\dots, w_{k}\}$).
\end{defin}
We can now state the following
\begin{thm}
Let $\Psi\in\mathbb{C}^{m}\otimes\mathbb{C}^{n}$ and $C(\Psi)$ denote the matrix of coefficients of $\Psi$ in an arbitrary common product basis. Then $\Psi$ is a product state if and only if $\dim\mathfrak{W}=1$ (or alternatively $\dim\mathfrak{V}=1$).
\end{thm}
\begin{proof}
Let $\{e_{i}\}_{i=1,\dots,m}$ and $\{f_{j}\}_{j=1,\dots,n}$ denote bases in $\mathbb{C}^{m}$ and $\mathbb{C}^{n}$, respectively. Then any vector $\Psi\in\mathbb{C}^{m}\otimes\mathbb{C}^{n}$ can be written as
\begin{equation}\label{GenExp2Dim}
\Psi=\sum_{i=1}^{m}\sum_{j=1}^{n}c_{ij}e_{i}\otimes f_{j},
\end{equation}
\noindent where $c_{ij}$ are the complex coeficients of $\Psi$, which we arrange as:
\begin{equation}
C(\Psi)\equiv\begin{pmatrix}
c_{11}&\dots&c_{1n}\\
\vdots&\ddots&\vdots\\
c_{m1}&\dots&c_{mn}
\end{pmatrix}.
\end{equation}
The matrix $C(\Psi)\equiv C$ always admits a SVD, given by $C=V\Sigma W^{\dagger}$, where $V$ and $W$ are unitary matrices and $\Sigma$ is a diagonal matrix with entries $\sigma_{k}$ (the singular values, indeed). Thus
\begin{equation}\label{SVDC}
c_{ij}=\sum_{k=1}^{\min(m,n)}v_{ik}\sigma_{k}w_{jk}^{*}.
\end{equation}
Inserting \eqref{SVDC} into \eqref{GenExp2Dim} and identifying new bases $\{\bar{e}_{i}\}_{i=1,2}$ and $\{\bar{f}_{j}\}_{j=1,2}$ we arrive at the well-known Schmidt decomposition
\begin{equation}\label{Schmidt}
\Psi=\sum_{k=1}^{\min(m,n)}\sigma_{k}\bar{e}_{k}\otimes\bar{f}_{k}.
\end{equation}
The number of non-null singular values coincides with the rank of $\Sigma$, which in turn coincides with the dimensions of $\mathfrak{V}$ and $\mathfrak{W}$ (cf.\ appendix).
\end{proof}
From the proof we can deduce a practical method to recognize where a bipartite system is entangled or not:
\begin{corol}
Let $\Psi\in\mathbb{C}^{m}\otimes\mathbb{C}^{n}$ denote the state of a bipartite quantum system and $C(\Psi)$ its coefficient matrix in an arbitrary product basis. Then $\Psi$ is a product state if, and only if, $\rg(C(\Psi))=1$.
\end{corol}
\subsection{Classification of two-qubit entanglement under SLOCC}
We only need one further tool to find the classification of bipartite entanglement under SLOCC, which is established as follows:
\begin{prop}
Let $\Psi,\bar{\Psi}\in\mathbb{C}^{2}\otimes\mathbb{C}^{2}$ denote two two-qubit states related by SLOCC, i.e.
\begin{equation}\label{NewPsi}
\bar{\Psi}=F^{[1]}\otimes F^{[2]}(\Psi),
\end{equation}
\noindent where $F^{[1]}$ and $F^{[2]}$ are non-singular operators upon $\mathbb{C}^{2}$. Then their corresponding coefficient matrices $C, \bar{C}$ in an arbitrary product basis are related through
\begin{equation}
\bar{C}=(F^{[1]^{T}}V)\Sigma (F^{[2]\dagger}W)^{\dagger}.
\end{equation}

\end{prop}

\begin{proof}
Just substitute $\Psi=\sum_{i,j=1,2}c_{ij}e_{i}\otimes f_{j}$ in \eqref{NewPsi} and identify indices.
\end{proof}
The key idea in our analysis is to recognize the effect of the ILO's $F^{[i]}$ upon the singular vectors. If $v_{j}$ (resp.\ $w_{j}$) is a left (resp.\ right) singular vector for the matrix coefficient $C$, then $F^{[1]T}(v_{j})$ (resp.\ $F^{[2]\dagger}(w_{j})$) is a left (resp.\ right) ``singular vector''\footnote{Notice that they cannot rigorously be  singular vectors, since the ILO's are not in general unitary, thus they do not preserve the orthogonality of $\{v_{j}\}$ and $\{w_j\}$. We will understand these ``singular vectors'' in a loose sense, in which they substitute the original singular vectors in the SVD of the coefficient matrix.} for the new matrix coefficient $\bar{C}$. In order to ease the notation, we will agree hereafter on relating $\Psi$ and $\bar{\Psi}$ through $\bar{\Psi}=F^{[1]T}\otimes F^{[2]\dagger}(\Psi)$, which allows us to drop the transpose and Hermitian conjugation\footnote{The transpose and Hermitian conjugation are referred to the chosen product basis in which $C$ is constructed.} in future considerations.\\
The case of two qubits is elementary, since there is no much space to discuss. The bases in which the coefficient matrix will be expressed are the canonical orthonormal basis $\{e_{1},e_{2}\}$ in $\mathbb{C}^{2}$. Only two options are present: either $\dim\mathfrak{W}=1$ or $\dim\mathfrak{W}=2$. In the first case, after choosing $F^{[1]}$ such that
\begin{subequations}
\begin{eqnarray}
F^{[1]}(v_{1})&=&\frac{1}{\sigma_{1}}e_{1},\\
F^{[2]}(w_{1})&=&e_{1},
\end{eqnarray}
\end{subequations}
\noindent the new coefficient matrix will turn into $\bar{C}=\left(\begin{smallmatrix}1&0\\0&0\end{smallmatrix}\right)$, which corresponds to the product state $\bar{\Psi}=e_{1}\otimes e_{1}$. We will agree on stating that $\Psi$ belongs to the entanglement class denoted by $00$.\\
In the second case, where $\sigma_{1}\geq\sigma_{2}>0$, after choosing $F^{[1]}$ and $F^{[2]}$ such that

\begin{subequations}
\begin{eqnarray}
F^{[1]}(v_{1})=\frac{1}{\sigma_{1}}e_{1},&\qquad& F^{[1]}(v_{2})=\frac{1}{\sigma_{2}}e_{2},\\
F^{[2]}(w_{1})=e_{1},&\qquad& F^{[2]}(w_{2})=e_{2},
\end{eqnarray}
\end{subequations}
\noindent the new coefficient matrix will be $\bar{C}=\left(\begin{smallmatrix}1&0\\0&1\end{smallmatrix}\right)$, which corresponds to the entangled state $\bar{\Psi}=e_{1}\otimes e_{1}+e_{2}\otimes e_{2}$. Now we say that $\Psi$ belongs to the class $\Psi^{+}$. \\
The reader can readily check by simple inspection how in the first case the canonical matrix $\bar{C}$ has rank one, whereas in the second it has rank $2$, as expected. In summary, only two classes are possible, namely $00$ and $\Psi^{+}$.
\section{Tripartite entanglement}
\label{TriEntSec}
The classification of tripartite pure states is performed along the same lines, namely choosing the ILO's $F^{[i]}$ so that the final coefficient matrix reduces to a canonical one. In order to find such canonical matrices, we must be exhaustive in the considerations of all possibilities when discussing about $\mathfrak{V}$ and $\mathfrak{W}$.\\
 The analysis of tripartite entanglement can be undertaken upon three possible coefficient matrices, arising from the three different ways to group the indices, that is, since $\Psi=\sum_{i_{1},i_{2},i_{3}=1,2}c_{i_{1}i_{2}i_{3}}e_{i_{1}}\otimes e_{i_{2}}\otimes e_{i_{3}}$, where as before $\{e_{k}\}$ denotes the canonical orthonormal basis in $\mathbb{C}^{2}$, we have
\begin{subequations}
\begin{eqnarray}\label{Part1}
C^{(1)}\equiv C_{1|23}&=&\begin{pmatrix}
c_{111}&c_{112}&c_{121}&c_{122}\\
c_{211}&c_{212}&c_{221}&c_{222}
\end{pmatrix},\\\label{Part2}
C^{(2)}\equiv C_{2|13}&=&\begin{pmatrix}
c_{111}&c_{112}&c_{211}&c_{212}\\
c_{121}&c_{122}&c_{221}&c_{222}
\end{pmatrix},\\\label{Part3}
C^{(3)}\equiv C_{3|12}&=&\begin{pmatrix}
c_{111}&c_{121}&c_{211}&c_{221}\\
c_{112}&c_{122}&c_{212}&c_{222}
\end{pmatrix}.
\end{eqnarray}
\end{subequations}
There is no loss of generality in choosing one of them, since the analysis will be exhaustive. Hereafter we will choose $C=C^{(1)}$. Notice that now the left singular vectors of $C$ belong to $\mathbb{C}^{2}$ whereas the right singular vectors are in $\mathbb{C}^{2}\otimes\mathbb{C}^{2}$. Also, we immediately realize that only two possibles options arise, namely $\dim\mathfrak{W}=1$ or $\dim\mathfrak{W}=2$, since there are at most two positive singular values. The recursivity appears when classifying the different structures which the subspace $\mathfrak{W}$ can show. The classification of these subspaces is performed according to the entanglement classes which their generators belong to. In order to do that we need the following result, which was firstly proved in the context of entanglement theory in \cite{SanTarVid98a}. We offer an alternative proof in order to illustrate our methods.
\begin{prop}\label{StruW}
Any two-dimensional subspace in $\mathbb{C}^{2}\otimes\mathbb{C}^{2}$ contains at least one product vector.
\end{prop}
\begin{proof}
Let $V$ be a two-dimensional subspace of $\mathbb{C}^{2}\otimes\mathbb{C}^{2}$. With no loss of generality two entangled vectors can be chosen as generators of $V$  with coefficient matrices given by $C_{1}=\mathbb{I}$ and $C_{2}$ being an arbitrary rank-2 matrix in the product canonical basis. Then it is always possible to find non-null complex numbers $\alpha$ and $\beta$ such that $\alpha\mathbb{I}+\beta C_{2}$ has rank one\footnote{Notice that $-\beta/\alpha$ must be chosen to be an eigenvalue of $C_{2}$.}.
\end{proof}
In other words, this proposition shows that $\lin\{\Psi_{1},\Psi_{2}\}$ always equals either $\lin\{\phi_{1}\otimes\psi_{1},\phi_{2}\otimes\psi_{2}\}$ or $\lin\{\phi\otimes\psi, \Psi\}$, where implicit are the assumptions that different indices denote linear independence and in the last case only one product unit vector can be found. Thus, with the same convention, the right singular subspace $\mathfrak{W}$ can show six different structures, namely $\lin\{\phi\otimes\psi\}$, $\lin\{\Psi\}$, $\lin\{\phi\otimes\psi_{1},\phi\otimes\psi_{2}\}$, $\lin\{\phi_{1}\otimes\psi,\phi_{2}\otimes\psi\}$, $\lin\{\phi_{1}\otimes\psi_{1},\phi_{2}\otimes\psi_{2}\}$ and  $\lin\{\phi\otimes\psi,\Psi\}$. We pursue Prop.\ \ref{StruW} a step further:
\begin{prop}\label{StruW2}
Let $\mathfrak{W}$ be a two-dimensional subspace in $\mathbb{C}^{2}\otimes\mathbb{C}^{2}$. Then  $\mathfrak{W}=\lin\{\phi\otimes\varphi,\Psi\}$ if, and only if, $\mathfrak{W}=\lin\{\phi\otimes\varphi,\phi\otimes\bar{\varphi}+\bar{\phi}\otimes\varphi\}$, where\hspace*{1mm}  $\bar{ }$ denotes linear independence.
\end{prop}
\begin{proof}
Suppose $\phi\otimes\varphi$ is the only product vector in $\mathfrak{W}$ (up to normalization factors). Its orthogonal vector in $\mathfrak{W}$ will be an entangled vector with coordinates in a product basis $\{\phi,\bar{\phi}\}\otimes\{\varphi,\bar{\varphi}\}$ given by $(0,\alpha_{12},\alpha_{21},\alpha_{22})$, i.e.\ it will be of the form $\phi\otimes\bar{\varphi}+\bar{\phi}\otimes\varphi+a\bar{\phi}\otimes\bar{\varphi}$, with $a\in\mathbb{C}$. Since $\phi\otimes\varphi$ must be the only product vector in $\mathfrak{W}$, it necessarily has to be $a=0$; otherwise could it always be possible to find $\alpha,\beta\in\mathbb{C}$ such that $\alpha\phi\otimes\varphi+\beta\left(\bar{\phi}\otimes\varphi+\phi\otimes\bar{\varphi}+a\bar{\phi}\otimes\bar{\varphi}\right)$ is another product vector ($\beta=a\alpha$).\\
Suppose now that $\mathfrak{W}=\{\phi\otimes\varphi,\phi\otimes\bar{\varphi}+\bar{\phi}\otimes\varphi\}$, then $\alpha\phi\otimes\varphi+\beta\left(\phi\otimes\bar{\varphi}+\bar{\phi}\otimes\varphi\right)$ is a product vector if, and only if, $\beta=0$, i.e.\ if it is the original $\phi\otimes\varphi$.
\end{proof}
We can now state our result, already contained in \cite{DurVidCir00a} with different criteria:
\begin{thm}\label{TriEnt}
Let $\Psi\in\mathbb{C}^{2}\otimes\mathbb{C}^{2}\otimes\mathbb{C}^{2}$ be the pure state of a tripartite system. Then $\Psi$ can be reduced through SLOCC to one of the following six states, which corresponds to the six possible entanglement classes, according to the following table:
\begin{widetext}
\begin{center}
\begin{tabular}{|c|c|c|c|}\hline
 \emph{\textbf{Class}} & \emph{\textbf{Canonical vector}} & \emph{\textbf{Canonical matrix}} & $\mathfrak{W}$\\\hline
$000$ & $e_{1}\otimes e_{1}\otimes e_{1}$ & $\begin{pmatrix}
1 & 0 & 0 & 0\\
0 & 0 & 0 & 0
\end{pmatrix}$ & $\lin\{\phi\otimes\psi\}$\\\hline
$0_{1}\Psi_{23}^{+}$ & $e_{1}\otimes e_{1}\otimes e_{1}+e_{1}\otimes e_{2}\otimes e_{2}$ & $\begin{pmatrix}
1 & 0 & 0 & 1\\
0 & 0 & 0 & 0
\end{pmatrix}$ & $\lin\{\Psi\}$\\\hline
$0_{2}\Psi_{13}^{+}$ & $e_{1}\otimes e_{1}\otimes e_{1}+e_{2}\otimes e_{1}\otimes e_{2}$ & $\begin{pmatrix}
1 & 0 & 0 & 0\\
0 & 1 & 0 & 0
\end{pmatrix}$ & $\phi\otimes\mathbb{C}^{2}$\\\hline
$0_{3}\Psi_{12}^{+}$ & $e_{1}\otimes e_{1}\otimes e_{1}+e_{2}\otimes e_{2}\otimes e_{1}$ & $\begin{pmatrix}
1 & 0 & 0 & 0\\
0 & 0 & 1 & 0
\end{pmatrix}$ & $\mathbb{C}^{2}\otimes\psi$\\\hline
$GHZ$ & $e_{1}\otimes e_{1}\otimes e_{1}+e_{2}\otimes e_{2}\otimes e_{2}$ & $\begin{pmatrix}
1 & 0 & 0 & 0\\
0 & 0 & 0 & 1
\end{pmatrix}$ & $\lin\{\phi_{1}\otimes\psi_{1},\phi_{2}\otimes\psi_{2}\}$\\\hline
$W$ & $e_{1}\otimes e_{1}\otimes e_{2}+e_{1}\otimes e_{2}\otimes e_{1}+e_{2}\otimes e_{1}\otimes e_{1}$ & $\begin{pmatrix}
0 & 1 & 1 & 0\\
1 & 0 & 0 & 0
\end{pmatrix}$ & $\lin\{\phi_{1}\otimes\psi_{1},\Psi\}$\\\hline
\end{tabular}
\end{center}
\end{widetext}
\end{thm}
\begin{proof}
 We discuss depending on $\mathfrak{W}$:
\begin{enumerate}
\item $\mathfrak{W}=\lin\{\phi\otimes\psi\}$. In this case, $w_{1}=\phi\otimes\psi$. Choose the ILO's $F^{[k]}$, $k=1,2,3$ so that
\begin{subequations}
\begin{eqnarray}
F^{[1]}(v_{1})&=&\frac{1}{\sigma_{1}}e_{1},\\
F^{[2]}(\phi)&=&e_{1},\\
F^{[3]}(\psi)&=&e_{1}.
\end{eqnarray}
\end{subequations}
Then the new coefficient matrix will be
\begin{eqnarray}
\bar{C}&=&\begin{pmatrix}
\frac{1}{\sigma_{1}} & \cdot\\
0 & \cdot
\end{pmatrix}\cdot\begin{pmatrix}
\sigma_{1} & 0 & 0 & 0\\
0 & 0 & 0 & 0
\end{pmatrix}\cdot
\begin{pmatrix}
1 & 0 & 0 & 0\\
\cdot & \cdot & \cdot & \cdot\\
\cdot & \cdot & \cdot & \cdot\\
\cdot & \cdot & \cdot & \cdot
\end{pmatrix}\nn\\
&=&\begin{pmatrix}
1 & 0 & 0 & 0\\
0 & 0 & 0 & 0
\end{pmatrix},
\end{eqnarray}
\noindent which corresponds to the state $e_{1}\otimes e_{1}\otimes e_{1}$, and where the dots $\cdot$ indicates the irrelevant character of that entry.
\item $\mathfrak{W}=\lin\{\Psi\}$. In this case $w_{1}=\phi_{1}\otimes\psi_{1}+\phi_{2}\otimes\psi_{2}$. Choose the ILO's so that

\begin{subequations}
\begin{eqnarray}
F^{[1]}(v_{1})=\frac{1}{\sigma_{1}}e_{1},&&\\
F^{[2]}(\phi_{1})=e_{1},&\qquad & F^{[2]}(\phi_{2})=e_{2},\\
F^{[3]}(\psi_{1})=e_{1},&\qquad & F^{[3]}(\psi_{2})=e_{2}.
\end{eqnarray}
\end{subequations}
Then the new coefficient matrix will be

\begin{eqnarray}
\bar{C}&=&\begin{pmatrix}
\frac{1}{\sigma_{1}} & \cdot\\
0 & \cdot
\end{pmatrix}\cdot\begin{pmatrix}
\sigma_{1} & 0 & 0 & 0\\
0 & 0 & 0 & 0
\end{pmatrix}\cdot
\begin{pmatrix}
1 & 0 & 0 & 1\\
\cdot & \cdot & \cdot & \cdot\\
\cdot & \cdot & \cdot & \cdot\\
\cdot & \cdot & \cdot & \cdot
\end{pmatrix}\nn\\
&=&\begin{pmatrix}
1 & 0 & 0 & 1\\
0 & 0 & 0 & 0
\end{pmatrix},
\end{eqnarray}
\noindent which corresponds to the state $e_{1}\otimes e_{1}\otimes e_{1}+e_{1}\otimes e_{2}\otimes e_{2}$.

\item $\mathfrak{W}=\phi\otimes\mathbb{C}^{2}=\lin\{\phi\otimes\psi_{1},\phi\otimes\psi_{2}\}$. In this case $w_{1}=\mu_{11}\phi\otimes\psi_{1}+\mu_{12}\phi\otimes\psi_{2}$ and  $w_{2}=\mu_{21}\phi\otimes\psi_{1}+\mu_{22}\phi\otimes\psi_{2}$, where the matrix $[\mu_{ij}]$ has rank $2$, since $w_{1}$ and $w_{2}$ are linear independent (orthonormal, indeed). Choose the ILO's so that

\begin{subequations}
\begin{eqnarray}
&F_{1}^{[1]}(v_{1})=\frac{1}{\sigma_{1}}e_{1},\qquad F_{1}^{[1]}(v_{2})=\frac{1}{\sigma_{2}}e_{2},&\\
&F^{[1]}_{2}=[F_{2}^{[1]}(e_{1})\ F_{2}^{[1]}(e_{2})]=[\mu_{ij}^{*}]^{-1},&\\
&F^{[1]}=F_{2}^{[1]}F_{1}^{[1]},&\\
&F^{[2]}(\phi)=e_{1},&\\
&F^{[3]}(\psi_{1})=e_{1},\qquad F^{[3]}(\psi_{2})=e_{2}.&
\end{eqnarray}
\end{subequations}
Then the new coefficient matrix will be

\begin{eqnarray}
\bar{C}&=&\begin{pmatrix}
\mu_{11}^{*}& \mu_{12}^{*}\\
\mu_{21}^{*}& \mu_{22}^{*}
\end{pmatrix}^{-1}\cdot\begin{pmatrix}
\frac{1}{\sigma_{1}} & 0\\
0 & \frac{1}{\sigma_{2}}
\end{pmatrix}\nonumber\\
&&\cdot\begin{pmatrix}
\sigma_{1} & 0 & 0 & 0\\
0 & \sigma_{2} & 0 & 0
\end{pmatrix}\cdot\begin{pmatrix}
\mu_{11}^{*} & \mu_{12}^{*} & 0 & 0\\
\mu_{21}^{*} & \mu_{22}^{*} & 0 & 0\\
\cdot & \cdot & \cdot & \cdot\\
\cdot & \cdot & \cdot & \cdot
\end{pmatrix}\nonumber\\
&=&\begin{pmatrix}
1 & 0 & 0 & 0\\
0 & 1 & 0 & 0
\end{pmatrix},
\end{eqnarray}
\noindent which corresponds to the state $e_{1}\otimes e_{1}\otimes e_{1}+e_{2}\otimes e_{1}\otimes e_{2}$.

\item $\mathfrak{W}=\mathbb{C}^{2}\otimes\psi=\lin\{\phi_{1}\otimes\psi,\phi_{2}\otimes\psi\}$. In this case $w_{1}=\mu_{11}\phi_{1}\otimes\psi+\mu_{12}\phi_{2}\otimes\psi$ and  $w_{2}=\mu_{21}\phi_{1}\otimes\psi+\mu_{22}\phi_{2}\otimes\psi$, where the matrix $[\mu_{ij}]$ has rank $2$, since $w_{1}$ and $w_{2}$ are linear independent (orthonormal, indeed). Choose the ILO's so that

\begin{subequations}
\begin{eqnarray}
&F_{1}^{[1]}(v_{1})=\frac{1}{\sigma_{1}}e_{1},\qquad F_{1}^{[1]}(v_{2})=\frac{1}{\sigma_{2}}e_{2},&\\
&F^{[1]}_{2}=[F_{2}^{[1]}(e_{1})\ F_{2}^{[1]}(e_{2})]=[\mu_{ij}^{*}]^{-1},&\\
&F^{[1]}=F_{2}^{[1]}F_{1}^{[1]},&\\
&F^{[2]}(\phi_{1})=e_{1},\qquad F^{[2]}(\phi_{2})=e_{2},&\\
&F^{[3]}(\psi)=e_{1}.&
\end{eqnarray}
\end{subequations}
Then the new coefficient matrix will be

\begin{eqnarray}
\bar{C}&=&\begin{pmatrix}
\mu_{11}^{*}& \mu_{12}^{*}\\
\mu_{21}^{*}& \mu_{22}^{*}
\end{pmatrix}^{-1}\cdot\begin{pmatrix}
\frac{1}{\sigma_{1}} & 0\\
0 & \frac{1}{\sigma_{2}}
\end{pmatrix}\nn\\
&&\cdot\begin{pmatrix}
\sigma_{1} & 0 & 0 & 0\\
0 & \sigma_{2} & 0 & 0
\end{pmatrix}\cdot
\begin{pmatrix}
\mu_{11}^{*} & 0 & \mu_{12}^{*} & 0 \\
\mu_{21}^{*} & 0 & \mu_{22}^{*} & 0 \\
\cdot & \cdot & \cdot & \cdot\\
\cdot & \cdot & \cdot & \cdot
\end{pmatrix}\nonumber\\
&=&\begin{pmatrix}
1 & 0 & 0 & 0\\
0 & 0 & 1 & 0
\end{pmatrix},
\end{eqnarray}
\noindent which corresponds to the state $e_{1}\otimes e_{1}\otimes e_{1}+e_{2}\otimes e_{2}\otimes e_{1}$.

\item $\mathfrak{W}=\lin\{\phi_{1}\otimes\psi_{1},\phi_{2}\otimes\psi_{2}\}$. In this case $w_{1}=\mu_{11}\phi_{1}\otimes\psi_{1}+\mu_{12}\phi_{2}\otimes\psi_{2}$ and  $w_{2}=\mu_{21}\phi_{1}\otimes\psi_{1}+\mu_{22}\phi_{2}\otimes\psi_{2}$, where the matrix $[\mu_{ij}]$ has rank $2$, since $w_{1}$ and $w_{2}$ are linear independent (orthonormal, indeed). Choose the ILO's so that

\begin{subequations}
\begin{eqnarray}
&F_{1}^{[1]}(v_{1})=\frac{1}{\sigma_{1}}e_{1},\qquad F_{1}^{[1]}(v_{2})=\frac{1}{\sigma_{2}}e_{2},&\\
&F^{[1]}_{2}=[F_{2}^{[1]}(e_{1})\ F_{2}^{[1]}(e_{2})]=[\mu_{ij}^{*}]^{-1},&\\
&F^{[1]}=F_{2}^{[1]}F_{1}^{[1]},&\\
&F^{[2]}(\phi_{1})=e_{1},\qquad F^{[2]}(\phi_{2})=e_{2},&\\
&F^{[3]}(\psi_{1})=e_{1},\qquad F^{[3]}(\psi_{2})=e_{2}.&
\end{eqnarray}
\end{subequations}
Then the new coefficient matrix will be

\begin{eqnarray}
\bar{C}&=&\begin{pmatrix}
\mu_{11}^{*}& \mu_{12}^{*}\\
\mu_{21}^{*}& \mu_{22}^{*}
\end{pmatrix}^{-1}\cdot\begin{pmatrix}
\frac{1}{\sigma_{1}} & 0\\
0 & \frac{1}{\sigma_{2}}
\end{pmatrix}\nn\\
&&\cdot\begin{pmatrix}
\sigma_{1} & 0 & 0 & 0\\
0 & \sigma_{2} & 0 & 0
\end{pmatrix}\cdot
\begin{pmatrix}
\mu_{11}^{*} & 0 & 0 & \mu_{12}^{*}\\
\mu_{21}^{*} & 0 & 0 & \mu_{21}^{*}\\
\cdot & \cdot & \cdot & \cdot\\
\cdot & \cdot & \cdot & \cdot
\end{pmatrix}=\nonumber\\
&=&\begin{pmatrix}
1 & 0 & 0 & 0\\
0 & 0 & 0 & 1
\end{pmatrix},
\end{eqnarray}
\noindent which corresponds to the state $e_{1}\otimes e_{1}\otimes e_{1}+e_{2}\otimes e_{2}\otimes e_{2}$.

\item $\mathfrak{W}=\lin\{\phi_{1}\otimes\psi_{1},\Psi\}$. In this notation, remember that implicit is the assumption that only one product unit vector can be found in $\mathfrak{W}$. In this case $\Psi$ can be chosen so that $\Psi=\phi_{1}\otimes\psi_{2} + \phi_{2}\otimes\psi_{1}$ (this is the statement in Prop.\ \ref{StruW2}). Thus the singular vectors can always be expressed as $w_{1}=\mu_{11}\left(\phi_{1}\otimes\psi_{2} + \phi_{2}\otimes\psi_{1}\right)+\mu_{12}\phi_{1}\otimes\psi_{1}$ and  $w_{2}=\mu_{21}\left(\phi_{1}\otimes\psi_{2}+\phi_{2}\otimes\psi_{1}\right)+\mu_{22}\phi_{1}\otimes\psi_{1}$, where the matrix $[\mu_{ij}]$ has rank $2$, since $w_{1}$ and $w_{2}$ are linear independent (orthonormal, indeed). Choose the ILO's so that

\begin{subequations}
\begin{eqnarray}
&F_{1}^{[1]}(v_{1})=\frac{1}{\sigma_{1}}e_{1},\qquad F_{1}^{[1]}(v_{2})=\frac{1}{\sigma_{2}}e_{2},&\\
&F^{[1]}_{2}=[F_{2}^{[1]}(e_{1})\ F_{2}^{[1]}(e_{2})]=[\mu_{ij}^{*}]^{-1},&\\
&F^{[1]}=F_{2}^{[1]}F_{1}^{[1]},&\\
&F^{[2]}(\phi_{1})=e_{1}\qquad F^{[2]}(\phi_{2})=e_{2},&\\
&F^{[3]}(\psi_{1})=e_{1}\qquad F^{[3]}(\psi_{2})=e_{2}.&
\end{eqnarray}
\end{subequations}
Then the new coefficient matrix will be

\begin{eqnarray}
\bar{C}&=&\begin{pmatrix}
\mu_{11}^{*}& \mu_{12}^{*}\\
\mu_{21}^{*}& \mu_{22}^{*}
\end{pmatrix}^{-1}\cdot\begin{pmatrix}
\frac{1}{\sigma_{1}} & 0\\
0 & \frac{1}{\sigma_{2}}
\end{pmatrix}\nn\\
&&\cdot\begin{pmatrix}
\sigma_{1} & 0 & 0 & 0\\
0 & \sigma_{2} & 0 & 0
\end{pmatrix}\cdot
\begin{pmatrix}
\mu_{12}^{*} & \mu_{11}^{*} & \mu_{11}^{*} & 0\\
\mu_{22}^{*} & \mu_{21}^{*} & \mu_{21}^{*} & 0\\
\cdot & \cdot & \cdot & \cdot\\
\cdot & \cdot & \cdot & \cdot
\end{pmatrix}\nonumber\\
&=&\begin{pmatrix}
0 & 1 & 1 & 0\\
1 & 0 & 0 & 0
\end{pmatrix},
\end{eqnarray}
\noindent which corresponds to the state $e_{1}\otimes e_{1}\otimes e_{2}+e_{1}\otimes e_{2}\otimes e_{1}+e_{2}\otimes e_{1}\otimes e_{1}$.

\end{enumerate}
Since there is no more options for the subspace $\mathfrak{W}$ we have already considered all possible alternatives.

\end{proof}
 In conclusion, we have found that there are six classes of entanglement, named after \cite{DurVidCir00a} as $000$, $0_{i_{1}}\Psi^{+}_{i_{2}i_{3}}$, $GHZ$ and $W$. The theorem also indicates how to practically classify a given state $\Psi$: compute the SVD of its coefficient matrix and elucidate the structure of $\lin\{w_{1},w_{2}\}$. We include a further proposition comprising the practical implementation of this result. We need to introduce the following definition.
\begin{defin}
Let $w_{j}=e_{1}\otimes w_{j1}+e_{2}\otimes w_{j2}\in\mathbb{C}^{2}\otimes\mathbb{C}^{2}$ be an arbitrary vector. We associate a two-dimensional matrix $W_{j}$ to $w_{j}$ by defining
\begin{equation}
W_{j}=[w_{j1}\ w_{j2}].
\end{equation}
\end{defin}
This definition will be mainly applied to the right singular vectors of the coefficient matrix $C$. As usual, the singular values of $C$ will be denoted by $\sigma_{k}$, in nonincreasing order, and $\sigma(A)$ denotes the spectrum of a matrix $A$. Our proposal to implement the preceding result is
\begin{thm}
Let $\Psi$ denote the pure state of a tripartite system and $C^{(i)}$ its coefficient matrix according to the partitions $i|jk$ (cf.\ \eqref{Part1}-\eqref{Part3}). Then
\begin{enumerate}
\item $\Psi$ belongs to the $000$ class if, and only if, $\rg(C^{(i)})=1$ for all $i=1,2,3$.
\item $\Psi$ belongs to the $0_{1}\Psi^{+}_{23}$ class if, and only if, $\rg(C^{(1)})=1$ and $\rg(C^{(k)})=2$ for $k=2,3$.
\item $\Psi$ belongs to the $0_{2}\Psi^{+}_{13}$ class if, and only if, $\rg(C^{(2)})=1$ and $\rg(C^{(k)})=2$ for $k=1,3$.
\item $\Psi$ belongs to the $0_{3}\Psi^{+}_{12}$ class if, and only if, $\rg(C^{(3)})=1$ and $\rg(C^{(k)})=2$ for $k=1,2$.
\item $\Psi$ belongs to the $GHZ$ class if, and only if, one of the following situations occurs:
\begin{itemize}
\item[i.]  $\rg(C^{(i)})=2$ for all $i=1,2,3$ and $\rg(W_{1})=\rg(W_{2})=1$.
\item[ii.] $\rg(C^{(i)})=2$ for all $i=1,2,3$, $\rg(W_{1})=2$, $\rg(W_{2})=1$ and $\sigma(W_{1}^{-1}W_{2})$ is non-degenerate.
\item[iii.] $\rg(C^{(i)})=2$ for all $i=1,2,3$, $\rg(W_{2})=2$, $\rg(W_{1})=1$ and $\sigma(W_{2}^{-1}W_{1})$ is non-degenerate.
\item[iv.] $\rg (C^{(i)})=2$ for all $i=1,2,3$, $\rg(W_{1})=2$, $\rg(W_{2})=2$ and $\sigma(W_{1}^{-1}W_{2})$ is non-degenerate.
\end{itemize}
\item $\Psi$ belongs to the $W$ class if, and only if, one of the following situations occurs:
\begin{itemize}
\item[i.] $\rg(C^{(i)})=2$ for all $i=1,2,3$, $\rg(W_{1})=2$, $\rg(W_{2})=1$ and $\sigma(W_{1}^{-1}W_{2})$ is degenerate.
\item[ii.] $\rg(C^{(i)})=2$ for all $i=1,2,3$, $\rg(W_{1})=1$, $\rg(W_{2})=2$ and $\sigma(W_{2}^{-1}W_{1})$ is degenerate.
\item[iii.] $\rg(C^{(i)})=2$ for all $i=1,2,3$, $\rg(W_{1})=2$, $\rg(W_{2})=2$ and $\sigma(W_{1}^{-1}W_{2})$ is degenerate.
\end{itemize}
\end{enumerate}
\end{thm}
\begin{proof}
We will exclusively concentrate upon the sufficiency, since the necessity directly follows from the canonical form of each class.\\
The first four cases are elementary, since it is a matter of detection of the vector which factorizes. The final two cases correspond to true tripartite entangled states. If $\rg(W_{k})=1$ for $k=1,2$, it is clear that there exist two product vectors belonging to $\mathfrak{W}$, thus $\Psi$ belongs to the $GHZ$ class. If $\rg(W_{1})=2$ and $\rg(W_{2})=1$ we need to check whether an ILO applied upon the first qubit can reduce the rank of the transformed $\bar{W}_{1}$. As it can be deduced from the preceding proofs, an ILO upon the first qubit amounts to constructing a linear combination between the two right singular vectors, which is equivalent to find new matrices $\bar{W}_{j}=F^{[1]}_{1j}W_{1}+F^{[1]}_{2j}W_{2}$, with $j=1,2$. If $\rg(W_{1})=2$, then by multiplying this expression to the left by $W_{1}^{-1}$, we have
\begin{equation}
F^{[1]}_{1j}\mathbb{I}+F^{[1]}_{2j}W^{-1}_{1}W_{2}.
\end{equation}
It is immediate to realize that it is possible to reduce the rank of $W_{1}$ to $1$ and to choose $F^{[1]}_{ij}$ such that $F^{[1]}$ is nonsingular provided the spectrum of $W^{-1}_{1}W_{2}$ is non-degenerate, in which case $\Psi$ belongs to the $GHZ$ class. If the spectrum is degenerate, thus both eigenvalues being null, no further reduction is possible and $\Psi$ belongs to the $W$ class. The symmetric case runs along parallel lines.\\
Finally if $\rg(W_{1})=\rg(W_{2})=2$, reasoning along similar lines if both eigenvalues of $W_{1}^{-1}W_{2}$ are equal, only one rank can be reduced keeping the nonsingularity of $F^{[1]}$ and $\Psi$ belongs again to the $W$ class, whereas if the eigenvalues are different, both ranks can be reduced to $1$ keeping the nonsingularity of $F^{[1]}$ and $\Psi$ belongs to the $GHZ$ class.
\end{proof}
As a final remark let us indicate how close, despite the apparent differences in the approach, our analysis runs parallel to that performed in \cite{DurVidCir00a}: the ranges of the reduced  density operators are indeed generated by the corresponding singular vectors, and the study of these ranges drove them and has driven us to the same final result. The change of method is motivated by the attempt to find a generalizable criterion not using entanglement measures specifically built upon the number of qubits of the system, such as the $3-$tangle \cite{CofKunWoo00a}. With this approach it is not necessary to consider at any stage the reduced density matrices and entanglement measures upon them. A strongly related approach can be found in \cite{CheChe06a,CheCheMei06a}.
\section{Generalizations ($N\geq 4$)}
\label{Gener}
The generalization of the preceding approach to pure states of arbitrary multipartite systems is two-folded. On one hand, the generalization to multiqubit states can be implemented inductively:
\begin{thm}
If the entanglement classes under SLOCC are known for $N$ qubits, the corresponding entanglement classes for $N+1$ qubits are also known.
\end{thm}
\begin{proof}
We proceed by induction. We have proved in preceding sections that this statement is true for $N=2$ and have explicitly found the entanglement classes for $N=3$. For a given $(N+1)$-qubit system, write the coefficient matrix $C_{1|2\cdots N+1}\equiv C$. Because of the induction hypothesis one knows in advance the classification of the right singular subspaces of $C$ according to $\mathfrak{W}=\lin\{\Psi_{i}\}$ if $\dim\mathfrak{W}=1$ and $\mathfrak{W}=\lin\{\Psi_{i},\Psi_{j}\}$ if $\dim\mathfrak{W}=2$, where each $\Psi_{i}$ and $\Psi_{j}$ belong to one (possibly the same) of the entanglement classes of $N$ qubits. Choose the ILO's $F^{[2]}\otimes\dots\otimes F^{[N+1]}$ so that the two first columns of $\bar{W}$ (the transformed right singular vectors) are expressed as linear combinations of the canonical vectors of the entanglement classes corresponding to the structure of $\mathfrak{W}$  and choose the ILO $F^{[1]}$ so that $\bar{V}\Sigma\bar{W}^{\dagger}$ drops out as many non-null entries as possible (typically $F^{[1]}$ will be the inverse of a rank$-2$ submatrix of $W^{\dagger}$). The result is the canonical matrix for an entanglement class of $N+1$ qubits.
\end{proof}
There is an important remark in the preceding inductive construction, already stated in \cite{DurVidCir00a} and explicitly shown in \cite{VerDehMooVer02a}: there will be a continuous range of states with a similar right singular subspace but with no ILO's connecting them. Let us illustrate this peculiar fact with an explicit example. When considering $4-$partite entanglement, there will exist $45$ a priori structures of the right singular subspace of the coefficient matrix, arising from $6$ possible one-dimensional right singular subspaces $\mathfrak{W}=\lin\{\Psi\}$, where $\Psi$ belongs to one of the six entanglement classes of $N=3$, times $4$ possible sites for the fourth added qubit, plus $21$ possible bidimensional right singular subspaces $\mathfrak{W}=\lin\{\Psi_{1},\Psi_{2}\}$, corresponding to the $\binom{6+2-1}{2}$ ways to choose the classes for $N=3$ which $\Psi_{1}$ and $\Psi_{2}$ belong to. An example will be $\mathfrak{W}=\lin\{000,GHZ\}$, with the already convention that only one product vector and no $0_{i}\Psi_{jk}$ belongs to $\mathfrak{W}$, i.e.\ $\mathfrak{W}=\lin\{\phi_{1}\otimes\varphi_{1}\otimes\psi_{1}, \phi_{2}\otimes\varphi_{2}\otimes\psi_{2}+\bar{\phi}_{2}\otimes\bar{\varphi}_{2}\otimes\bar{\psi}_{2}\}$, where the vectors with\hspace*{1mm} $\bar{}$ are pairwise linearly independent. In order to only have one product vector and the rest being $GHZ$ vectors, we must have \footnote{A detailed account of the classification of $4-$qubit entanglement using this method is under preparation and will appear elsewhere.} (up to permutations) $\mathfrak{W}=\lin\{\phi\otimes\bar{\varphi}\otimes\psi', \phi\otimes\varphi\otimes\psi+\bar{\phi}\otimes\bar{\varphi}\otimes\bar{\psi}\}$, with $\psi'\neq\psi,\bar{\psi}$.\\
Recalling that
\begin{subequations}
\begin{eqnarray}
w_{1}&=&\mu_{11}\phi\otimes\bar{\varphi}\otimes\psi'+\mu_{12}\left(\phi\otimes\varphi\otimes\psi+\bar{\phi}\otimes\bar{\varphi}\otimes\bar{\psi}\right),\nn\\
&&\\
w_{2}&=&\mu_{21}\phi\otimes\bar{\varphi}\otimes\psi'+\mu_{22}\left(\phi\otimes\varphi\otimes\psi+\bar{\phi}\otimes\bar{\varphi}\otimes\bar{\psi}\right),\nn\\
\end{eqnarray}
\end{subequations}
\noindent where the matrix $[\mu_{ij}]\equiv\left(\begin{smallmatrix}\mu_{11}& \mu_{12}\\\mu_{21}&\mu_{22}\end{smallmatrix}\right)$ will be non-singular, it is immediate to find ILO's $F^{[2]}, F^{[3]}, F^{[4]}$ such that
\begin{widetext}
\begin{subequations}
\begin{eqnarray}
F^{[2]}\otimes F^{[3]}\otimes F^{[4]}(w_{1})&=&\mu_{11}e_{1}\otimes e_{2}\otimes\psi+\mu_{12}\left(e_{1}\otimes e_{1}\otimes e_{1}+e_{2}\otimes e_{2}\otimes e_{2}\right),\\
F^{[2]}\otimes F^{[3]}\otimes F^{[4]}(w_{2})&=&\mu_{21}e_{1}\otimes e_{2}\otimes\psi+\mu_{22}\left(e_{1}\otimes e_{1}\otimes e_{1}+e_{2}\otimes e_{2}\otimes e_{2}\right)
\end{eqnarray}
\end{subequations}
\end{widetext}
\noindent which corresponds to a coefficient matrix given by

\begin{equation}
\bar{C}=\bar{V}\Sigma\begin{pmatrix}
\mu_{12}^{*} & 0 & \mu_{11}^{*}\psi_{1}^{*} & \mu_{11}^{*}\psi_{2}^{*} & 0 & 0 & 0 & \mu_{12}^{*}\\
\mu_{22}^{*} & 0 & \mu_{21}^{*}\psi_{1}^{*} & \mu_{21}^{*}\psi_{2}^{*} & 0 & 0 & 0 & \mu_{22}^{*}\\
\cdot & \cdot &\cdot &\cdot &\cdot &\cdot &\cdot &\cdot \\
\vdots &\vdots &\vdots &\vdots &\vdots &\vdots &\vdots &\vdots \\
\cdot &\cdot &\cdot &\cdot &\cdot &\cdot &\cdot &\cdot
\end{pmatrix}_{8\times 8},
\end{equation}
\noindent where the coefficients $\psi_{i}$ corresponds to the coordinates of the transformed $\psi'$ in the canonical basis. Choosing $F^{[1]}$ so that
\begin{equation}
\bar{V}\Sigma=[\mu_{ij}^{*}]^{-1},
\end{equation}
\noindent we arrive at
\begin{equation}
\bar{C}=\begin{pmatrix}
0 & 0 & \psi_{1}^{*} & \psi_{2}^{*} & 0 & 0 & 0 & 0\\
1 & 0 & 0 & 0 & 0 & 0 & 0 & 1
\end{pmatrix},
\end{equation}
\noindent which corresponds to the canonical vector
\begin{widetext}
\begin{eqnarray}
e_{1}\otimes e_{1}\otimes e_{2}\otimes\psi^*+e_{2}\otimes e_{1}\otimes e_{1}\otimes e_{1}+e_{2}\otimes e_{2}\otimes e_{2}\otimes e_{2}=\quad (\psi^*\neq e_{1},e_{2})\nn\\
=\ket{001\psi^*}+\ket{1000}+\ket{1111}\quad (\ket{\psi^{*}}\neq \ket{0},\ket{1})\label{4ParCanVec}
\end{eqnarray}
\end{widetext}
Thus, different $\psi$ will yield different entanglement classes under non-singular local operators $F^{[1]}\otimes\dots\otimes F^{[N]}$. Notice that this vector belongs neither to the $GHZ_{4}$ class nor to the $W_{4}$ class nor to the $\Phi_{4}$ class (containing the cluster state of four qubits -see below). It is a peculiar feature that two infinitesimally close states could belong to distinct entanglement classes, so a deeper elucidation of this point is on due and will be carried out also elsewhere. For the time being, we will agree on attributing all states reducible to \eqref{4ParCanVec} by ILO's $F^{[1]}\otimes\dots\otimes F^{[4]}$, independently of the particular vector $\psi$, the same entanglement properties under SLOCC and analogously for arbitrary $N$-partite multiqubit systems.\\
This allows us to find an upper bound for the number of genuine $(N+1)$-partite entanglement classes. Firstly, notice that e.g.\ the right singular subspace $\mathfrak{W}=\lin\{000,000\}$ in the $4-$partite case actually contains structures with different properties, namely\footnote{As usual, different indices denote linear independence.} $\mathfrak{W}=\phi\otimes\varphi\otimes\mathbb{C}^{2}$ (and permutations), $\mathfrak{W}=\lin\{\phi\otimes\varphi_{1}\otimes\psi_{1},\phi\otimes\varphi_{2}\otimes\psi_{2}\}$ (and permutations) and  $\mathfrak{W}=\lin\{\phi_{1}\otimes\varphi_{1}\otimes\psi_{1},\phi_{2}\otimes\varphi_{2}\otimes\psi_{2}\}$. All of them drives us to at least one factor qubit in the final canonical state, except one, that is, there will correspond one right singular subspace structure $\lin\{\Psi_{1},\Psi_{2}\}$ to each genuine $(N+1)-$entanglement class.\\

This is rigorously proved in the following
\begin{prop}
Let $\mathfrak{W}_{N}$ be the right singular subspace of the coefficient matrix in an arbitrary product basis of an $N$-qubit pure state. If $\mathfrak{W}_{N}$ is supported in a product space $\mathfrak{W}_{N}=\psi\otimes\mathfrak{W}_{N-1}$, then the state belongs to a product class $0_{2}\Psi$, where $\Psi$ denotes a class of $(N-1)$-partite entanglement.
\end{prop}
\begin{proof}
Under the above assumption, $w_{j}=\psi\otimes\bar{w}_{j}$, $j=1,2$, with $\psi\in\mathbb{C}^{2}$ and $\bar{w}_{j}\in\mathbb{C}^{2(N-2)}$. We can always find an ILO $F^{[2]}$ such that
\begin{eqnarray}
\bar{w}_{j}&\to& e_{1}\otimes\hat{w}_{j},
\end{eqnarray}
\noindent where also  $\hat{w}_{j}\in\mathbb{C}^{2(N-2)}$, hence $\bar{W}_{N}=E_{11}\otimes\bar{W}_{N-1}$, where $E_{11}$ denotes the Weyl matrix $E_{11}=|e_{1}\rangle\langle e_{1}|$. Since we can always write $\Sigma_{N}=E_{11}\otimes\Sigma_{N-1}$, the coefficient matrix can always be written as
\begin{eqnarray}
\bar{C}_{N}&=&\bar{V}\Sigma_{N}\bar{W}^{\dagger}=\bar{V}\left(E_{11}\otimes\Sigma_{N-1}\right)\left(E_{11}\otimes\bar{W}_{N-1}\right)^{\dagger}\nn\\
&=&E_{11}\otimes\left(\bar{V}\Sigma_{N-1}\bar{W}_{N-1}^{\dagger}\right).
\end{eqnarray}
The remaining ILO's $F^{[1]}$ and $F^{[j]}$, $j>2$, can always be chosen so that
\begin{equation}
\bar{C}_{N}=E_{11}\otimes\bar{C}_{N-1},
\end{equation}
\noindent where $\bar{C}_{N-1}$ denotes a canonical matrix of an $(N-1)$-partite entanglement class. This proves that the second qubit factorizes, as the reader may check.
\end{proof}
With appropiate permutations, this result applies to any qubit. If we denote by $M(N)$ the number of $N-$partite entanglement classes, there will be at most
\begin{equation}
\binom{M(N)+2-1}{2}=\frac{1}{2}\left[M(N)+1\right]M(N)
\end{equation}
\noindent genuine entanglement classes for $N+1$ qubits. Besides, the number of degenerate $(N+1)-$entanglement classes will be at most $(N+1)\times M(N)$ (corresponding to the $N+1$ possible factor positions which the $(N+1)$th qubit can occupy), thus
\begin{corol}
Let $M(N)$ denote the number of $N-$partite entanglement classes under SLOCC. Then
\begin{equation}
M(N+1)\leq\frac{1}{2}M(N)\left[M(N)+2N+3\right]\label{UpperBound}.
\end{equation}
\end{corol}
 The equality will be in general unattainable, since, as in the case of tripartite entanglement, only a few distinct true entanglement classes exist, coming out from the only actually different structures which the right singular subspace can adopt (only two in the case of tripartite systems; cf.\ proposition \ref{StruW}).\\
Let us call reader's attention on the fact that these results allow us to view all state space of $N$ qubits divided into blocks, each one parametrized by a right singular subspace structure and corresponding to our broad-sense entanglement classes, and within which the difference between states stems from a (possibly several) continuous parameter. The number of these blocks for $N+1$ qubits is upperly bounded by the recursive relation \eqref{UpperBound}.\\
Another benefit of the present approach arises when deciding whether two states belong to the same entanglement class or not. This is stated as a corollary:
\begin{corol}
Let $\Psi,\Phi\in(\mathbb{C}^{2})^{\otimes N}$. Let $\mathfrak{W}_{\Psi}$ and $\mathfrak{W}_{\Phi}$ be their respective right singular subspaces. Then a necessary and sufficient condition for $\Psi,\Phi$ to belong to the same broad-sense entanglement class under SLOCC is that $\mathfrak{W}_{\Psi}$ and $\mathfrak{W}_{\Phi}$ have the same structure, i.e.\ that they are generated by entanglement-equivalent vectors.
\end{corol}
\begin{proof}
The result follows immediately both from construction and from the convention on the definition of the broad-sense entanglement classes.
\end{proof}
 As an example, let us include a one-line proof that the $4-$qubit GHZ state $|GHZ_{4}\rangle\equiv\frac{1}{\sqrt{2}}\left(|0000\rangle+|1111\rangle\right)$ and the cluster state $|\phi_{4}\rangle\equiv\frac{1}{2}\left(|0000\rangle+|0011\rangle+|1100\rangle-|1111\rangle\right)$ \cite{BriRau01a} do not belong to the same class \cite{WuZha00a}. Their respective right singular subspaces are $\mathfrak{W}_{GHZ_4}=\lin\{e_{1}\otimes e_{1}\otimes e_{1},e_{2}\otimes e_{2}\otimes e_{2}\}$ and $\mathfrak{W}_{\phi_{4}}=\lin\{e_{1}\otimes\Psi^{+},e_{2}\otimes\Psi^{-}\}$, where $\Psi^{\pm}$ denote two-qubit Bell states. It is immediate to conclude that they are different, since none $e_{j}\otimes e_{j}\otimes e_{j}$ belong to $\mathfrak{W}_{\phi_{4}}$ (write the coefficient matrix of a generic vector in $\mathfrak{W}_{\phi_{4}}$ in terms of two coordinates $\alpha$ and $\beta$ and check that it is impossible to choose the latter so that the matrix corresponds to $e_{j}\otimes e_{j}\otimes e_{j}$). These states belong to the respective so-called\footnote{The first one is named by a natural extension of the tripartite case; the second, after its representative $|\phi_{4}\rangle$.} $GHZ_{4}$ and $\Phi_{4}$ classes, characterized by the above right singular subspaces.\\
On the other hand, to find a wider generalization one can focus upon arbitrary dimensional entangled systems. The \emph{leit motiv} is still the same, with the important exception that the dimension of the right singular subspace can grow up to the dimension of the Hilbert space of the first subsystem. Thus the analysis of the possible structures which $\mathfrak{W}$ may adopt is now much more complex. \\
We include as an illustrative immediate example the analysis of all entanglement classes under SLOCC of any bipartite $(N_{1}\times N_{2})$-dimensional system: there exist $\min(N_{1},N_{2})$ entanglement classes, which can be denoted as $00\equiv\Psi_{1}^+$, $\Psi_{2}^{+}$, $\Psi_{3}^{+}$, \dots, $\Psi_{\min(N_{1}, N_{2})}^{+}$, whose canonical states will elementarily be $\sum_{i=1}^{k}e_{i}\otimes e_{i}$, for each class $\Psi_{k}^{+}$. They correspond to canonical matrices given by $\sum_{i=1}^{k}\ket{e_{i}^{[N_{1}]}}\bra{e_{i}^{[N_{2}]}}$, so that we can state the following
\begin{thm}
Let $\Psi\in\mathbb{C}^{N_{1}}\otimes\mathbb{C}^{N_{2}}$ be the pure state of a bipartite quantum system with coefficient matrix in an arbitrary product basis denoted by $C(\Psi)$. Then $\Psi$ belongs to the $\Psi_{k}^{+}$ class, $k=1,2,\dots,\min(N_{1}, N_{2})$, if, and only if, $\rg(C(\Psi))=k=\dim\mathfrak{V}=\dim\mathfrak{W}$.
\end{thm}
\begin{proof}
Let $\mathfrak{V}=\lin\{\phi_{k}\}_{k=1,\cdots,n\leq\min(N_{1},N_{2})}$ and  $\mathfrak{W}=\lin\{\varphi_{k}\}_{k=1,\cdots,n\leq\min(N_{1},N_{2})}$. Choose $F^{[1]}$ and $F^{[2]}$ so that
\begin{eqnarray}
F^{[1]}(\phi_{k})&=&\frac{1}{\sigma_{k}}e_{k},\\
F^{[2]}(\varphi_{k})&=&e_{k}.
\end{eqnarray}
Then the coefficient matrix (in blocks) will turn out to be
\begin{equation}
\bar{C}=\begin{pmatrix}
\mathbb{I}_{n} & 0_{N_{2}-n}\\
0_{N_{1}-n} & 0_{N_{1}-n,N_{2}-n}
\end{pmatrix}.
\end{equation}
\end{proof}
For more general cases, the difference stems solely in the higher computational complexity.
\section{Conclusions}
\label{Con}
We have developed a recursive inductive criterion to classify entanglement under SLOCC in multipartite systems in pure states which allows one to find the entanglement classes for $N+1$ qubits provided this classification is known for $N$ qubits. The method rests on the analysis of the right singular subspace of their coefficient matrix, which is chosen according to the partition $1|2\dots N$, hence a $2\times 2^{N-1}$ rectangular matrix. Then one must elucidate the classification of the one- and two-dimensional right singular subspaces according to the entanglement classes which their generators belong to.  As a consequence, this construction reveals a systematic way to detect the entanglement class of a given state without resorting to quantitative measures of entanglement. In arbitrary-dimensional generalizations, the same scheme must be followed with the exception that the dimension of the right singular subspaces is higher and their structure now depends on several generators.\\
 For $N\geq 4$ it has been showed that within each right singular subspace structure, there could exist a continuous infinity of states not connected through invertible local operators. Additionally, up to this continuous degree of freedom within each right singular subspace structure, we have found an upper bound for the number of classes on $N+1$ qubits in terms of the number of classes of $N$ qubits.\\
As a final remark, let us conjecture that a possible connection with the MPS formalism is probable to exist. In this formalism (cf.\ \cite{Eck05a} and multiple references therein) any pure state is written as $$\Psi=\sum_{i_{1}\dots i_{N}}\textrm{tr}\left(A^{[i_{1}]}_{1}\dots A^{[i_{N}]}_{N}\right)e_{i_{1}}\otimes\dots\otimes e_{i_{N}},$$\noindent so that adjoining a further $(N+1)$-th qubit amounts to adjoining a further $A^{[i_{N+1}]}_{N+1}$ matrix in the trace giving the coefficients. In the analysis carried out above, this last added qubit is equivalent to increase the dimension of the right singular subspace $\dim\mathfrak{W}_{N}\to\dim\mathfrak{W}_{N+1}=2\times\dim\mathfrak{W}_{N}$. Our conjecture is that the structure of $\mathfrak{W}_{N}$ should be read from the properties of the $N$ matrices $A^{[i_{k}]}_{k}$, so that the succesion of structures of $\mathfrak{W}_{N}$ should run parallel to that of the matrices $A^{[i_{1}]}_{1},\dots, A^{[i_{N}]}_{N}$.

\appendix
\section{The singular value decomposition}
\label{AppSVD}
We include the relevant properties of the SVD of an arbitrary matrix and suggest the interested reader to consult e.g.\ \cite{HorJoh91a} for a comprehensive analysis of this decomposition with the corresponding proofs. The set of $m\times n$ complex matrices will be denoted as usual by $\mathcal{M}_{m,n}(\mathbb{C})\equiv\mathcal{M}_{m,n}$ and the group of unitary matrices of dimension $k$ will be denoted by $U(k)$. The main result can be stated as
\begin{thm} \textbf{(Singular Value Decomposition)}
Let $Q\in\mathcal{M}_{m,n}$. Then $Q$ can always be decomposed as
\begin{equation}
Q=V\Sigma W^{\dagger},
\end{equation}
\noindent where $V\in U(m)$, $W\in U(n)$ and $\Sigma\in M_{m,n}$ is a diagonal matrix with non-negative entries, i.e. $\Sigma_{ij}=\sigma_{i}\delta_{ij}$, with $i=1,\dots,m$,  $j=1,\dots,n$ and $\sigma_{k}\geq 0$ for all $k$.
\end{thm}
The columns of $V$ and $W$ and the positive entries of $\Sigma$ receive a special name:
\begin{defin}
The columns of $V=[v_{1}\ v_{2}\ \dots\ v_{m}]$ (resp.\ $W=[w_{1}\ w_{2}\ \dots\ w_{n}]$) are the left (resp.\ right) singular vectors of $Q$. The positive entries of $\Sigma$ are the singular values of $Q$.
\end{defin}
Notice that with this definition any $m\times n$ dimensional matrix will have $m$ left singular vectors and $n$ right singular vectors; since the relevant singular vectors will be those associated to non-null singular values, we agree, as usual, on referring as singular vectors only to the latter, i.e. to those $v_{k}$ and $w_{k}$ for which $\sigma_{k}>0$. Another common convention is the decreasing order of the singular values in the diagonal of $\Sigma$: $\sigma_{1}\geq\sigma_{2}\geq\dots\geq 0$.\\
The singular vectors are highly nonunique or equivalently there always exist another unitary matrices $\hat{V}$ and $\hat{W}$ such that $Q=\hat{V}\Sigma\hat{W}^{\dagger}$, where these new unitary matrices depend of the former $V$ and $W$ and the multiplicities of each singular value \cite{HorJoh91a}. However this fact has not been exploited in the text.\\
One of the main consequences of the SVD is that the rank of a given matrix $Q$ coincides with the rank of $\Sigma$, i.e.\ with the number of positive singular values, which, in turn, coincides with the dimension of the subspace generated by the left (or right) singular vectors. This is the basis to the analysis of entanglement of a pure state upon its coefficient matrix in a product basis performed in the text.
\section*{Acknowledgments}
The authors acknowledge financial support from the Spanish MEC
projects No.\ FIS2005-05304 (L.L. and J.L.) and No.\ FIS2004-01576
(D.S.) and from EU RESQ, EuroSQIP, and DFG SFB631 projects (E.S.).
L.L.\ also acknowledges support from the FPU grant No.\ AP2003-0014.

\begin{thebibliography}{30}
\expandafter\ifx\csname natexlab\endcsname\relax\def\natexlab#1{#1}\fi
\expandafter\ifx\csname bibnamefont\endcsname\relax
  \def\bibnamefont#1{#1}\fi
\expandafter\ifx\csname bibfnamefont\endcsname\relax
  \def\bibfnamefont#1{#1}\fi
\expandafter\ifx\csname citenamefont\endcsname\relax
  \def\citenamefont#1{#1}\fi
\expandafter\ifx\csname url\endcsname\relax
  \def\url#1{\texttt{#1}}\fi
\expandafter\ifx\csname urlprefix\endcsname\relax\def\urlprefix{URL }\fi
\providecommand{\bibinfo}[2]{#2}
\providecommand{\eprint}[2][]{\url{#2}}
\bibitem[{\citenamefont{Peres}(1993)}]{Per93a}
\bibinfo{author}{\bibfnamefont{A.}~\bibnamefont{Peres}},
  \emph{\bibinfo{title}{{Q}uantum {T}heory: {C}oncepts and {M}ethods}}
  (\bibinfo{publisher}{Kluwer}, \bibinfo{address}{Dordrecht},
  \bibinfo{year}{1993}).
\bibitem[{\citenamefont{Bennett and DiVincenzo}(2000)}]{BenDiV00a}
\bibinfo{author}{\bibfnamefont{C.H.}~\bibnamefont{Bennett}} \bibnamefont{and}
  \bibinfo{author}{\bibfnamefont{D.}~\bibnamefont{DiVincenzo}},
  \bibinfo{journal}{Nature} \textbf{\bibinfo{volume}{404}},
  \bibinfo{pages}{247} (\bibinfo{year}{2000}).
\bibitem[{\citenamefont{Dusek et~al.}()\citenamefont{Dusek, Lutkenhaus, and
  Hendrych}}]{DusLutHen06a}
\bibinfo{author}{\bibfnamefont{M.}~\bibnamefont{Dusek}},
  \bibinfo{author}{\bibfnamefont{N.}~\bibnamefont{Lutkenhaus}},
  \bibnamefont{and} \bibinfo{author}{\bibfnamefont{M.}~\bibnamefont{Hendrych}},
  \bibinfo{journal}{To appear in E. Wolf (ed.) Progress in Optics, vol. 49}
  (2006).
\bibitem[{\citenamefont{Deutsch and Ekert}(1998)}]{DeuEke98a}
\bibinfo{author}{\bibfnamefont{D.}~\bibnamefont{Deutsch}} \bibnamefont{and}
  \bibinfo{author}{\bibfnamefont{A.}~\bibnamefont{Ekert}},
  \bibinfo{journal}{Phys. World} \textbf{\bibinfo{volume}{11}},
  \bibinfo{pages}{47} (\bibinfo{year}{1998}).
\bibitem[{\citenamefont{Raussendorf and Briegel}(2001)}]{RauBri01a}
\bibinfo{author}{\bibfnamefont{R.}~\bibnamefont{Raussendorf}} \bibnamefont{and}
  \bibinfo{author}{\bibfnamefont{H.}~\bibnamefont{Briegel}},
  \bibinfo{journal}{Phys. Rev. Lett.} \textbf{\bibinfo{volume}{86}},
  \bibinfo{pages}{5188} (\bibinfo{year}{2001}).
\bibitem[{\citenamefont{Bouwmeester et~al.}(2000)\citenamefont{Bouwmeester,
  Ekert, and Zeilinger}}]{BouEkeZei00a}
\bibinfo{editor}{\bibfnamefont{D.}~\bibnamefont{Bouwmeester}},
  \bibinfo{editor}{\bibfnamefont{A.}~\bibnamefont{Ekert}}, \bibnamefont{and}
  \bibinfo{editor}{\bibfnamefont{A.}~\bibnamefont{Zeilinger}}, eds.,
  \emph{\bibinfo{title}{{T}he {P}hysics of {Q}uantum {I}nformation}}
  (\bibinfo{publisher}{Springer}, \bibinfo{address}{Berlin},
  \bibinfo{year}{2000}).
\bibitem[{\citenamefont{Horodecki et~al.}(2001)\citenamefont{Horodecki,
  Horodecki, and Horodecki}}]{HorHorHor01a}
\bibinfo{author}{\bibfnamefont{M.}~\bibnamefont{Horodecki}},
  \bibinfo{author}{\bibfnamefont{P.}~\bibnamefont{Horodecki}},
  \bibnamefont{and}
  \bibinfo{author}{\bibfnamefont{R.}~\bibnamefont{Horodecki}},
  \emph{\bibinfo{title}{Quantum Information: An Introduction to Basic
  Theoretical Concepts and Experiments}} (\bibinfo{publisher}{Springer},
  \bibinfo{address}{Berlin}, \bibinfo{year}{2001}), chap.
  \bibinfo{chapter}{Mixed-state entanglement and quantum communication},
  Springer Tracts in Modern Physics.
\bibitem[{\citenamefont{Bennett and Wiesner}(1992)}]{BenWie92a}
\bibinfo{author}{\bibfnamefont{C.H.}~\bibnamefont{Bennett}} \bibnamefont{and}
  \bibinfo{author}{\bibfnamefont{S.}~\bibnamefont{Wiesner}},
  \bibinfo{journal}{Phys. Rev. Lett.} \textbf{\bibinfo{volume}{69}},
  \bibinfo{pages}{2881} (\bibinfo{year}{1992}).
\bibitem[{\citenamefont{Bell}(1987)}]{Bel87a}
\bibinfo{author}{\bibfnamefont{J.}~\bibnamefont{Bell}},
  \emph{\bibinfo{title}{{S}peakable and {U}nspeakable in {Q}uantum
  {M}echanics}} (\bibinfo{publisher}{Cambridge University Press},
  \bibinfo{address}{Cambridge}, \bibinfo{year}{1987}).

\bibitem{EisGro05a}
\bibinfo{author}{\bibnamefont{J.}~\bibnamefont{Eisert}} \bibnamefont{and}
\bibinfo{authot}{\bibnamefont{D.}~\bibnamefont{Gross}},
  \bibinfo{journal}{quant-ph/0505149}.

\bibitem[{\citenamefont{Bru$\beta$}(2002)}]{Bru02a}
\bibinfo{author}{\bibfnamefont{D.}~\bibnamefont{Bru$\beta$}},
  \bibinfo{journal}{J. Math. Phys.} \textbf{\bibinfo{volume}{43}},
  \bibinfo{pages}{4237} (\bibinfo{year}{2002}).
\bibitem[{\citenamefont{D\"{u}r et~al.}(2000)\citenamefont{D\"{u}r, Vidal, and
  Cirac}}]{DurVidCir00a}
\bibinfo{author}{\bibfnamefont{W.}~\bibnamefont{D\"{u}r}},
  \bibinfo{author}{\bibfnamefont{G.}~\bibnamefont{Vidal}}, \bibnamefont{and}
  \bibinfo{author}{\bibfnamefont{J.I.}~\bibnamefont{Cirac}},
  \bibinfo{journal}{Phys. Rev. A} \textbf{\bibinfo{volume}{62}},
  \bibinfo{pages}{062314} (\bibinfo{year}{2000}).
\bibitem[{\citenamefont{Verstraete et~al.}(2002)\citenamefont{Verstraete,
  Dehaene, Moor, and Verschelde}}]{VerDehMooVer02a}
\bibinfo{author}{\bibfnamefont{F.}~\bibnamefont{Verstraete}},
  \bibinfo{author}{\bibfnamefont{J.}~\bibnamefont{Dehaene}},
  \bibinfo{author}{\bibfnamefont{B.} \bibnamefont{De Moor}}, \bibnamefont{and}
  \bibinfo{author}{\bibfnamefont{H.}~\bibnamefont{Verschelde}},
  \bibinfo{journal}{Phys. Rev. A} \textbf{\bibinfo{volume}{65}},
  \bibinfo{pages}{052112} (\bibinfo{year}{2002}).
\bibitem[{\citenamefont{Coffman et~al.}(2000)\citenamefont{Coffman, Kundu, and
  Wooters}}]{CofKunWoo00a}
\bibinfo{author}{\bibfnamefont{V.}~\bibnamefont{Coffman}},
  \bibinfo{author}{\bibfnamefont{J.}~\bibnamefont{Kundu}}, \bibnamefont{and}
  \bibinfo{author}{\bibfnamefont{W.}~\bibnamefont{Wooters}},
  \bibinfo{journal}{Phys. Rev. A} \textbf{\bibinfo{volume}{61}},
  \bibinfo{pages}{052306} (\bibinfo{year}{2000}).
\bibitem[{\citenamefont{Verstraete et~al.}(2003)\citenamefont{Verstraete,
  Dehaene, and Moor}}]{VerDehMoo03a}
\bibinfo{author}{\bibfnamefont{F.}~\bibnamefont{Verstraete}},
  \bibinfo{author}{\bibfnamefont{J.}~\bibnamefont{Dehaene}}, \bibnamefont{and}
  \bibinfo{author}{\bibfnamefont{B.} \bibnamefont{De Moor}},
  \bibinfo{journal}{Phys. Rev. A} \textbf{\bibinfo{volume}{68}},
  \bibinfo{pages}{012103} (\bibinfo{year}{2003}).
\bibitem[{\citenamefont{Vidal}(2000)}]{Vid00a}
\bibinfo{author}{\bibfnamefont{G.}~\bibnamefont{Vidal}}, \bibinfo{journal}{J.
  Mod. Opt.} \textbf{\bibinfo{volume}{47}}, \bibinfo{pages}{355}
  (\bibinfo{year}{2000}).

\bibitem{OstSie05a}
\bibinfo{author}{\bibfnamefont{A.}~\bibnamefont{Osterloch}},
  \bibnamefont{and} \bibinfo{author}{\bibfnamefont{J.}~\bibnamefont{Siewert}},
  \bibinfo{journal}{Phys. Rev. A} \textbf{\bibinfo{volume}{72}},
  \bibinfo{pages}{012337} (\bibinfo{year}{2005}).
\bibitem{LovMaaSmiAmiGraIliIzmZag06a}
 \bibinfo{author}{\bibfnamefont{P.J.}~\bibnamefont{Love}},
 \bibinfo{author}{\bibfnamefont{A.}~\bibnamefont{Maasen van den Brink}},
 \bibinfo{author}{\bibfnamefont{A.\ Yu.}~\bibnamefont{Smirnov}},
 \bibinfo{author}{\bibfnamefont{M.H.S.}~\bibnamefont{Amin}},
 \bibinfo{author}{\bibfnamefont{M.}~\bibnamefont{Grajcar}},
 \bibinfo{author}{\bibfnamefont{E.}~\bibnamefont{Il'ichev}},
 \bibinfo{author}{\bibfnamefont{A.}~\bibnamefont{Izmalkov}},
 \bibnamefont{and} \bibinfo{author}{\bibnamefont{A.M.}~\bibnamefont{Zagoskin}},
  \bibinfo{journal}{quant-ph/0602143}.

\bibitem{RigOliOli06a}
 \bibinfo{author}{\bibfnamefont{G.}~\bibnamefont{Rigolin}},
 \bibinfo{author}{\bibfnamefont{T.R.}~\bibnamefont{de Oliveira}},
 \bibnamefont{and} \bibinfo{author}{\bibnamefont{M.C.}~\bibnamefont{de Oliveira}},
  \bibinfo{journal}{quant-ph/0603215}.
\bibitem[{\citenamefont{Grassl et~al.}(1998)\citenamefont{Grassl, R\"{o}tteler,
  and Beth}}]{GraRotBet98a}
\bibinfo{author}{\bibfnamefont{M.}~\bibnamefont{Grassl}},
  \bibinfo{author}{\bibfnamefont{M.}~\bibnamefont{R\"{o}tteler}},
  \bibnamefont{and} \bibinfo{author}{\bibfnamefont{T.}~\bibnamefont{Beth}},
  \bibinfo{journal}{Phys. Rev. A} \textbf{\bibinfo{volume}{58}},
  \bibinfo{pages}{1833} (\bibinfo{year}{1998}).
\bibitem[{\citenamefont{Ac\'{\i}n et~al.}(2000)\citenamefont{Ac\'{\i}n,
  Andrianov, Costa, Jan\'{e}, Latorre, and Tarrach}}]{AciAndCosJanLatTar00a}
\bibinfo{author}{\bibfnamefont{A.}~\bibnamefont{Ac\'{\i}n}},
  \bibinfo{author}{\bibfnamefont{A.}~\bibnamefont{Andrianov}},
  \bibinfo{author}{\bibfnamefont{L.}~\bibnamefont{Costa}},
  \bibinfo{author}{\bibfnamefont{E.}~\bibnamefont{Jan\'{e}}},
  \bibinfo{author}{\bibfnamefont{J.I.}~\bibnamefont{Latorre}}, \bibnamefont{and}
  \bibinfo{author}{\bibfnamefont{R.}~\bibnamefont{Tarrach}},
  \bibinfo{journal}{Phys. Rev. Lett.} \textbf{\bibinfo{volume}{85}},
  \bibinfo{pages}{1560} (\bibinfo{year}{2000}).
\bibitem[{\citenamefont{Carteret et~al.}(2002)\citenamefont{Carteret, Higuchi,
  and Sudbery}}]{CarHigSud02a}
\bibinfo{author}{\bibfnamefont{H.}~\bibnamefont{Carteret}},
  \bibinfo{author}{\bibfnamefont{A.}~\bibnamefont{Higuchi}}, \bibnamefont{and}
  \bibinfo{author}{\bibfnamefont{A.}~\bibnamefont{Sudbery}},
  \bibinfo{journal}{J. Math. Phys.} \textbf{\bibinfo{volume}{41}},
  \bibinfo{pages}{7932} (\bibinfo{year}{2002}).
\bibitem[{\citenamefont{Gao et~al.}(2006)\citenamefont{Gao, Albeverio, Fei, and
  Wang}}]{GaoAlbFeiWan06a}
\bibinfo{author}{\bibfnamefont{X.}~\bibnamefont{Gao}},
  \bibinfo{author}{\bibfnamefont{S.}~\bibnamefont{Albeverio}},
  \bibinfo{author}{\bibfnamefont{S.}~\bibnamefont{Fei}}, \bibnamefont{and}
  \bibinfo{author}{\bibfnamefont{Z.}~\bibnamefont{Wang}},
  \bibinfo{journal}{Commun. Theor. Phys.} \textbf{\bibinfo{volume}{45}},
  \bibinfo{pages}{267} (\bibinfo{year}{2006}).

\bibitem[{\citenamefont{Eckholt}(2005)}]{Eck05a}
\bibinfo{author}{\bibfnamefont{M.}~\bibnamefont{Eckholt}}, Master's thesis,
  \bibinfo{school}{Technische Universit\"{a}t M\"{u}nchen/Max-Planck-Institut
  f\"{u}r Quantenoptik}, \bibinfo{address}{Garching} (\bibinfo{year}{2005}).
\bibitem[{\citenamefont{Affleck et~al.}(1987)\citenamefont{Affleck, Kennedy,
  Lieb, and Tasaki}}]{AffKenLieTas87a}
\bibinfo{author}{\bibfnamefont{I.}~\bibnamefont{Affleck}},
  \bibinfo{author}{\bibfnamefont{T.}~\bibnamefont{Kennedy}},
  \bibinfo{author}{\bibfnamefont{E.}~\bibnamefont{Lieb}}, \bibnamefont{and}
  \bibinfo{author}{\bibfnamefont{H.}~\bibnamefont{Tasaki}},
  \bibinfo{journal}{Phys. Rev. Lett.} \textbf{\bibinfo{volume}{59}},
  \bibinfo{pages}{799} (\bibinfo{year}{1987}).
\bibitem[{\citenamefont{Vidal}(2003)}]{Vid03a}
\bibinfo{author}{\bibfnamefont{G.}~\bibnamefont{Vidal}},
  \bibinfo{journal}{Phys. Rev. Lett.} \textbf{\bibinfo{volume}{91}},
  \bibinfo{pages}{147902} (\bibinfo{year}{2003}).
\bibitem[{\citenamefont{Verstraete et~al.}(2004)\citenamefont{Verstraete,
  Porras, and Cirac}}]{VerPorCir04a}
\bibinfo{author}{\bibfnamefont{F.}~\bibnamefont{Verstraete}},
  \bibinfo{author}{\bibfnamefont{D.}~\bibnamefont{Porras}}, \bibnamefont{and}
  \bibinfo{author}{\bibfnamefont{J.I.}~\bibnamefont{Cirac}},
  \bibinfo{journal}{Phys. Rev. Lett.} \textbf{\bibinfo{volume}{93}},
  \bibinfo{pages}{227205} (\bibinfo{year}{2004}).
\bibitem[{\citenamefont{Sch\"{o}n et~al.}(2005)\citenamefont{Sch\"{o}n, Solano,
  Verstraete, Cirac, and Wolf}}]{SchSolVerCirWol05a}
\bibinfo{author}{\bibfnamefont{C.}~\bibnamefont{Sch\"{o}n}},
  \bibinfo{author}{\bibfnamefont{E.}~\bibnamefont{Solano}},
  \bibinfo{author}{\bibfnamefont{F.}~\bibnamefont{Verstraete}},
  \bibinfo{author}{\bibfnamefont{J.I.}~\bibnamefont{Cirac}}, \bibnamefont{and}
  \bibinfo{author}{\bibfnamefont{M.}~\bibnamefont{Wolf}},
  \bibinfo{journal}{Phys. Rev. Lett.} \textbf{\bibinfo{volume}{95}},
  \bibinfo{pages}{110503} (\bibinfo{year}{2005}).
\bibitem[{\citenamefont{Schmidt}(1907)}]{Sch07a}
\bibinfo{author}{\bibfnamefont{E.}~\bibnamefont{Schmidt}},
  \bibinfo{journal}{Math. Ann.} \textbf{\bibinfo{volume}{63}},
  \bibinfo{pages}{433} (\bibinfo{year}{1907}).
\bibitem[{\citenamefont{Ekert and Knight}(1995)}]{EkeKni95a}
\bibinfo{author}{\bibfnamefont{A.}~\bibnamefont{Ekert}} \bibnamefont{and}
  \bibinfo{author}{\bibfnamefont{P.L.}~\bibnamefont{Knight}},
  \bibinfo{journal}{Am. J. Phys.} \textbf{\bibinfo{volume}{63}},
  \bibinfo{pages}{415} (\bibinfo{year}{1995}).
\bibitem[{\citenamefont{Sanpera et~al.}(1998)\citenamefont{Sanpera, Tarrach,
  and Vidal}}]{SanTarVid98a}
\bibinfo{author}{\bibfnamefont{A.}~\bibnamefont{Sanpera}},
  \bibinfo{author}{\bibfnamefont{R.}~\bibnamefont{Tarrach}}, \bibnamefont{and}
  \bibinfo{author}{\bibfnamefont{G.}~\bibnamefont{Vidal}},
  \bibinfo{journal}{Phys. Rev. A} \textbf{\bibinfo{volume}{58}},
  \bibinfo{pages}{826} (\bibinfo{year}{1998}).

\bibitem{CheChe06a}
\bibinfo{author}{\bibfnamefont{L.}~\bibnamefont{Chen}},
  \bibnamefont{and} \bibinfo{author}{\bibfnamefont{Y.-X.}~\bibnamefont{Chen}},
  \bibinfo{journal}{Phys. Rev. A} \textbf{\bibinfo{volume}{73}},
  \bibinfo{pages}{052310} (\bibinfo{year}{2006}).

\bibitem{CheCheMei06a}
\bibinfo{author}{\bibfnamefont{L.}~\bibnamefont{Chen}},
  \bibinfo{author}{\bibfnamefont{Y.-X.}~\bibnamefont{Chen}},
  \bibnamefont{and} \bibinfo{author}{\bibnamefont{Y.-X.}~\bibnamefont{Mei}},
  \bibinfo{journal}{quant-ph/0604184}.
\bibitem[{\citenamefont{Briegel and Raussendorf}(2005)}]{BriRau01a}
\bibinfo{author}{\bibfnamefont{H.}~\bibnamefont{Briegel}} \bibnamefont{and}
  \bibinfo{author}{\bibfnamefont{R.}~\bibnamefont{Raussendorf}},
  \bibinfo{journal}{Phys. Rev. Lett.} \textbf{\bibinfo{volume}{86}},
  \bibinfo{pages}{910} (\bibinfo{year}{2005}).
\bibitem[{\citenamefont{Wu and Zhang}(2000)}]{WuZha00a}
\bibinfo{author}{\bibfnamefont{S.}~\bibnamefont{Wu}} \bibnamefont{and}
  \bibinfo{author}{\bibfnamefont{Y.}~\bibnamefont{Zhang}},
  \bibinfo{journal}{Phys. Rev. A} \textbf{\bibinfo{volume}{63}},
  \bibinfo{pages}{012308} (\bibinfo{year}{2000}).
\bibitem[{\citenamefont{Horn and Johnson}(1991)}]{HorJoh91a}
\bibinfo{author}{\bibfnamefont{R.}~\bibnamefont{Horn}} \bibnamefont{and}
  \bibinfo{author}{\bibfnamefont{C.}~\bibnamefont{Johnson}},
  \emph{\bibinfo{title}{{T}opics in {M}atrix {A}nalysis}}
  (\bibinfo{publisher}{Cambridge University Press},
  \bibinfo{address}{Cambridge}, \bibinfo{year}{1991}).
\end{thebibliography}

\end{document}